\documentclass{article} 
\usepackage[final]{colm2025_conference}

\usepackage{microtype}
\usepackage{hyperref}
\usepackage{url}
\usepackage{booktabs}
\usepackage{tabularx}
\usepackage{lineno}

\usepackage{amsthm}

\usepackage{graphicx}
\usepackage{subcaption}
\usepackage{float}
\usepackage{multirow}

\usepackage{amsthm}  
\usepackage{ragged2e} 

\usepackage{newtxtext,newtxmath}

\newtheorem{theorem}{Theorem}[section] 
\newtheorem{proposition}[theorem]{Proposition} 
\newtheorem{definition}[theorem]{Definition} 
\newtheorem{assumption}[theorem]{Assumption} 
\newtheorem{property}[theorem]{Property} 

\title{The Economics of Information Pollution in the Age of AI: General Equilibrium, Welfare, and Policy Design}

\author{Yukun Zhang \\
The Chinese University Of Hongkong\\
HongKong, China \\
\texttt{215010026@link.cuhk.edu.cn} \\
\And
Tianyang Zhang \\
University of Bologna \\
Bologna, Italy \\
\texttt{tianyang.zhang@studio.unibo.it} \\
}

\begin{document}

\ifcolmsubmission
\linenumbers
\fi

\maketitle
\begin{abstract}
The rapid diffusion of generative artificial intelligence (AI) is triggering structural change in information markets, with impacts far more complex than simple 'technological progress'. This paper is the first to model AI as an \textbf{asymmetric technological shock}: in low-quality content production, AI acts as a \textbf{substitute} for labor ($\sigma_L > 1$), whereas in high-quality creation, it functions merely as a \textbf{complement} to human expertise ($\sigma_H < 1$). This technological asymmetry fundamentally alters the cost structure of information production, systematically favoring 'lemons' over 'peaches'. By constructing a three-stage general equilibrium model, we prove this shock leads to an inefficient \textbf{'Polluted Information Equilibrium'} sustained by a triad of interacting market failures: (1) a \textbf{Production Externality}, as low-quality producers do not internalize ecological harm; (2) a \textbf{Platform Governance Failure}, as engagement-based revenue models misalign algorithms with social welfare; and (3) a \textbf{Trust Commons Externality}, as verification, a public good, is systematically under-provided. To address this, we design an adaptive governance framework, first deriving a theoretically-grounded \textbf{Information Pollution Index (IPI)} for real-time ecosystem monitoring. Second, we demonstrate that restoring efficiency requires a \textbf{multi-instrument policy portfolio}: a Pigouvian tax to correct the production externality, content provenance standards to address under-verification, and fiduciary duties to constrain platform behavior. Finally, through Agent-Based Model (ABM) validation, we find this portfolio generates \textbf{superadditive welfare gains}—the joint intervention is significantly more effective than any single policy tool. The core insight is that the welfare consequences of the AI revolution depend not on the technology itself, but on how we design market rules; without proper governance, technological progress can paradoxically reduce social welfare, a phenomenon we term the \textbf{'Paradox of AI Progress'}.
\end{abstract}

\section{Introduction}
\label{sec:intro}

In 2025, an in-depth investigative report, the product of a senior journalist's three-week effort, garners 100,000 views on social media. On the same day, a sensational but unverified "scoop" generated by an AI in three seconds receives 10 million shares. This stark contrast reveals the fundamental challenge facing the information economy in the age of generative AI: when the marginal cost of producing convincing but unverified content approaches zero, how can truth compete with noise?

This is not merely a technical question; it is a core economic one. The rise of large language models (LLMs) like GPT-4 and Claude represents a structural shift in the information production function. However, existing research has largely focused on AI's impact on the labor market, overlooking a more fundamental problem: how does AI alter the equilibrium distribution of \textit{information quality}?

Consider two distinct content production scenarios:

\begin{itemize}
    \item \textbf{Scenario A (Investigative Report):} A journalist uses AI tools to accelerate data analysis and assist in literature review, but the core work—field interviews, fact-checking, and ethical judgment—still requires human expertise. Here, AI is a \textit{complement}, not a substitute.

    \item \textbf{Scenario B (Content Farm):} An operator uses AI to batch-generate plausible health advice, investment "insider" tips, or political "revelations". No professional knowledge is needed, only prompt engineering. Here, AI is a \textit{substitute}, almost entirely replacing human labor.
\end{itemize}

This asymmetry in production technology—what we term "technology-biased information pollution"—is reshaping the digital information ecosystem. By some estimates, 90\% of online content may be AI-generated or AI-assisted by 2026. If the majority of this is unverified, low-quality content, we face not just "information overload," but systemic "information pollution"—a new class of market failure.

The central thesis of this paper is that generative AI is not a neutral productivity tool, but rather a \textit{biased technological shock} that systematically lowers the production cost of "lemons" (low-quality content) relative to "peaches" (high-quality content). We argue that this technological bias leads to market failure through three mutually reinforcing mechanisms: at the \textit{micro-level}, individual producers rationally choose to produce low-quality content due to its low cost and rapid, algorithm-driven rewards; at the \textit{meso-level}, platform algorithms designed to maximize engagement amplify content based on virality rather than veracity; and at the \textit{macro-level}, a systemic erosion of trust leads to "verification fatigue" among consumers, further degrading the market's ability to discern quality.

This "pollution spiral," if left unchecked, risks a "tragedy of the digital commons," where everyone is a victim, yet no single actor has the incentive to unilaterally change behavior.

This paper's contributions are threefold: a theoretical innovation that is the first to model AI's market impact as an asymmetric Constant Elasticity of Substitution (CES) production function, revealing a paradox of progress; a measurement tool that constructs a multi-dimensional Information Pollution Index (IPI) to provide a quantitative basis for policy intervention; and a policy design that proves the necessity of a multi-instrument portfolio, validated through computational simulation.

The remainder of this paper is organized as follows. Section \ref{sec:lit_review} reviews the relevant literature and situates our contribution. Section \ref{sec:theory} builds the theoretical model and characterizes the equilibrium. Section \ref{sec:ipi} derives the IPI and designs the optimal policy portfolio. Section \ref{sec:policy_abm} provides validation via an Agent-Based Model (ABM). Section \ref{sec:conclusion} concludes and discusses policy implications.

\begin{figure*}[t]
\centering
\includegraphics[width=0.9\textwidth]{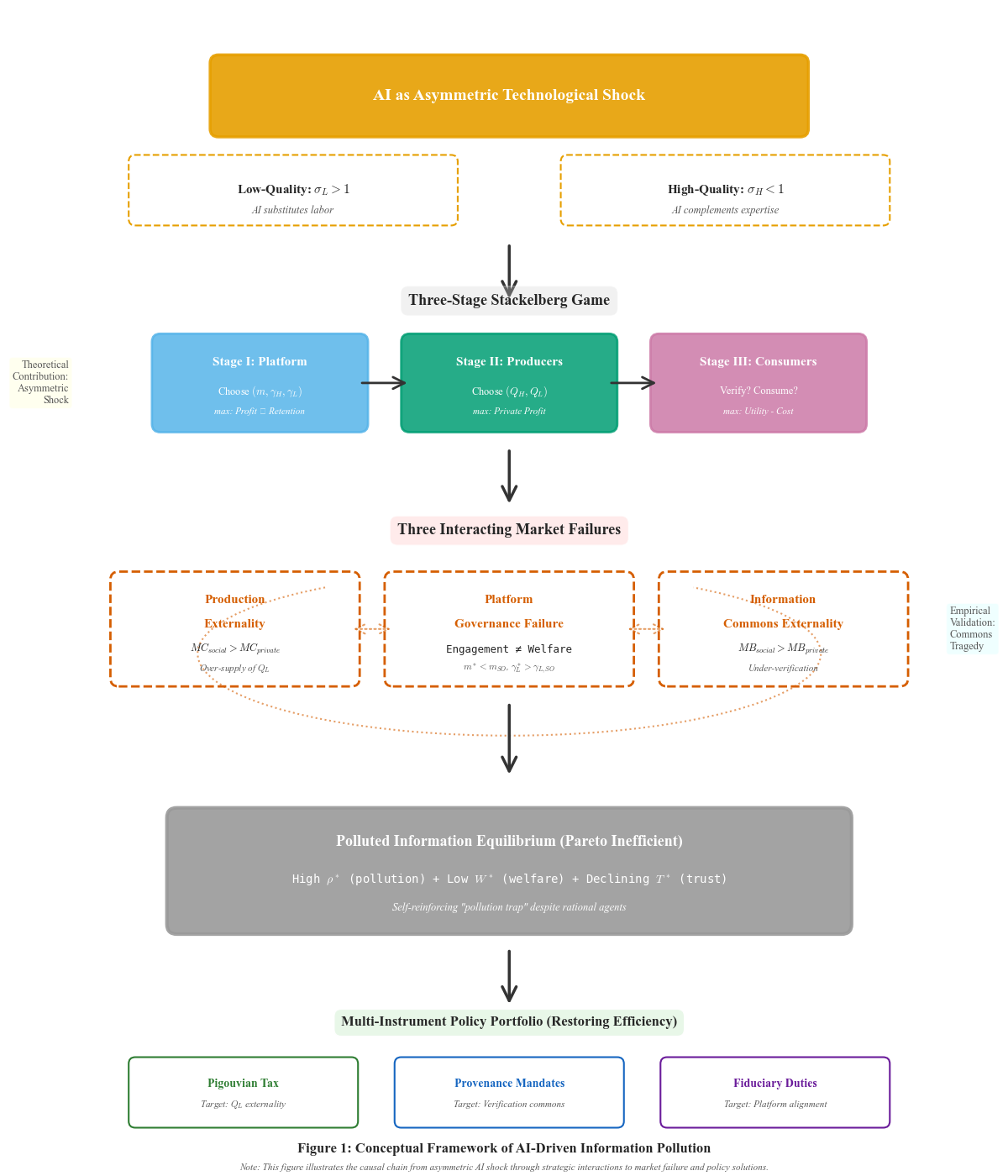} 
\caption{Conceptual Framework of AI-Driven Information Pollution. 
The model proceeds from the core asymmetric shock ($\sigma_L > 1 > \sigma_H$) through a three-stage Stackelberg game, which leads to a triad of interacting market failures: a Production Externality, a Platform Governance Failure, and an Information Commons Externality. These failures produce a self-reinforcing, Pareto-inefficient "Polluted Information Equilibrium." We show this equilibrium necessitates a multi-instrument policy portfolio (Pigouvian Tax, Provenance Mandates, Fiduciary Duties) to restore efficiency.}
\label{fig:conceptual_framework}
\end{figure*}

\section{Literature Review}
\label{sec:lit_review}

\subsection{Economic Foundations and Traditional Governance of Information}
The novel market failure of information pollution builds on Akerlof’s foundational ``market for lemons'' model \cite{Akerlof1970}, but manifests in far more complex, multi-dimensional ways in today’s digital environment. From the standpoint of economic theory, environmental economics provides two classic policy instruments for managing pollution: Pigouvian taxes and tradable permits.

Posner and Weyl \cite{Posner2018}, in their book \textit{Radical Markets}, explicitly advocated for taxing online advertising. Cabral et al. \cite{Cabral2021} further developed this idea and proposed a differentiation mechanism for low-quality content. In Zuboff’s \cite{Zuboff2019} book \textit{The Age of Surveillance Capitalism}, her profound critique of technology companies’ massive acquisition of user data to predict and change human behaviour provides an important theoretical basis and justification for regulating data exploitation and information pollution through economic means (such as taxation).

The tradable permit system is a mature governance tool in environmental economics. The core of this mechanism is setting a strict cap on total pollution emissions (Cap) and allocating emission rights as a market-tradable asset. Its theoretical foundation stems from Coase’s \cite{Coase1960} property rights theory and was first systematically proposed by Dales \cite{Dales1968}. Subsequently, Montgomery’s \cite{Montgomery1972} research theoretically demonstrated that, under specific conditions, this mechanism can achieve predetermined emission reduction targets with the lowest total social cost. Although there are some obstacles to shifting this perspective to information pollution, some scholars have put forward some relevant proposals.

In addition, some scholars have proposed market-based solutions from the perspective of platform governance. For example, the research of Parker and Van Alstyne \cite{Parker2010} inspired using reputation systems or information quality points to incentivise platforms to solve information asymmetry. Additionally, Crémer et al. \cite{Cremer2019} proposed mandatory data sharing and interoperability – essentially ``access quotas'' for core resources, similar to tradable Licences – to break monopolies and promote competition.

\subsection{AI, Algorithmic Amplification, and Ecosystem Degradation}

Beyond economics, scholars in information science and communication studies have emphasised the ecological nature of digital knowledge systems. Floridi \cite{Floridi2010} conceptualises information environments as ``infospheres'', arguing that pollution arises when the semantic integrity of these ecosystems is undermined. Bawden and Robinson \cite{Bawden2009} describe the resulting cognitive and social fatigue as a form of ``information environmental degradation''. These perspectives reinforce the economic analogy: information pollution, like environmental pollution, depletes a scarce common resource – trust.

Recent literature underscores that the information environment has entered a phase where quantity and veracity are inversely correlated. False news spreads faster and farther than verified information \cite{Vosoughi2018}, amplified by algorithmic incentives that optimise for engagement rather than truth \cite{Allcott2017,allcott2019trends}. This phenomenon is continuously reinforced by cognitive biases (such as the lazy thinking effect revealed by Pennycook and Rand \cite{Pennycook2019}) and causes multidimensional economic distortions, such as those highlighted by Allcott et al. \cite{Allcott2020a}, Durante et al. \cite{Durante2019}, and Bursztyn et al. \cite{Bursztyn2020}.

Algorithmic amplification further intensifies these dynamics. As Tufekci \cite{Tufekci2015} observes, recommendation systems can generate systemic harms even without malicious intent. Empirical evidence confirms that algorithmic curation narrows informational diversity \cite{Bakshy2015}. From a systems perspective, Helbing \cite{Helbing2018} frames these feedback loops as emergent properties of self-organising social systems, suggesting that agent-based modelling can illuminate how micro-level incentives produce macro-level information distortions.

This paper contributes to and bridges three major strands of literature. First, it extends the classic literature on information economics in the context of digital content, exploring how AI undermines traditional signalling mechanisms \cite{Spence1973}. The erosion of credibility parallels earlier discussions on asymmetric information and adverse selection, but now operates within algorithmically mediated attention markets. Second, it advances research in platform economics and industrial organisation \cite{Hagiu2015,rochet2003platform}. By integrating AI-driven production asymmetry into a two-sided platform framework, we provide a formal treatment of how content generation technologies interact with intermediation and market power. Third, it contributes to the emerging field of AI economics \cite{Acemoglu2019,Agrawal2018}, shifting the focus from labour displacement to the quality of informational output in one of society’s most critical domains – the information market.

\section{Theoretical Framework}
\label{sec:theory}

\subsection{Conceptual Foundation: Information as a Common-Pool Resource}
\label{sec:conceptual}

\subsubsection{The Credible Information Commons}
\label{sec:commons}

We conceptualize the ecosystem of credible information as a common-pool resource (CPR) in the Ostromian sense, rivalrous in quality yet only imperfectly excludable in access. Let the stock of high-quality information at time $t$, denoted by $S_t$, represent the attention-weighted reservoir of credible content circulating across digital platforms. New content creation replenishes this reservoir through a regenerative flow $R_t$, while low-quality or misleading information generated by actors seeking visibility, clicks, or advertising rents degrades it at a pollution rate $P_t$. The dynamic evolution of the resource follows
\begin{equation}
    \dot{S}_t = R_t - P_t(S_t,\,a_t,\,\theta_t),
\end{equation}
where $a_t$ captures the aggregate extraction intensity of attention by content producers, and $\theta_t$ summarizes governance and verification institutions that regulate entry, provenance, and moderation. The credible-information commons is renewable but exhaustible: its replenishment depends on costly verification and sustained trust, while degradation arises from unpriced externalities of generative AI and algorithmic amplification.

Extraction in this domain refers to the appropriation of limited audience attention $A_t$ for private content visibility. A simple representation of allocation shares is $s_i = \frac{q_i A_t}{\sum_j q_j A_t}$. Pollution corresponds to the dissemination of low-credibility or synthetic material that reduces the expected information accuracy $\mathbb{E}[h_i]$ and undermines collective trust $T_t = f(S_t)$. The marginal social cost of pollution, $\partial C^{soc}/\partial P_t$, is convex in $P_t$ because reputational erosion and verification fatigue accelerate once misinformation exceeds cognitive processing capacity.

To move from the tragedy to sustainable governance, Ostrom’s design principles can be adapted to digital information. Boundaries become traceable provenance and authentication of content and accounts, which can be parameterized by the verifiability intensity $m$. Congruence between rules and local conditions corresponds to aligning platform recommendation objectives $(\gamma_H, \gamma_L)$ with societal welfare by rewarding credible engagement over sheer volume. Collective choice arrangements imply participatory moderation and community fact-checking that internalize positive monitoring externalities. Monitoring and graduated sanctions suggest layered detection and proportionate penalties for repeated offenders, representable as increasing marginal cost of pollution $c_P(P_t)$. Conflict-resolution mechanisms provide low-cost correction channels to restore truthful information without discouraging legitimate debate. Finally, nested governance combines platform-level rules with national and international regulatory oversight, which we capture by multi-layer institutional variables $\theta_t = (\theta^{plat}, \theta^{nat}, \theta^{intl})$.

\subsubsection{The Tragedy of AI-Driven Pollution}
\label{sec:tragedy}

Each producer chooses output $q_i$ and quality $h_i$ to maximize private returns,
\begin{equation}
    \pi_i = p(h_i,q_i) - c(h_i,q_i),
\end{equation}
where $p(\cdot)$ is the monetized visibility price determined by platform algorithms. Because platforms typically reward engagement rather than veracity, producers internalize revenue from exposure but not the social cost of degraded credibility. In aggregate, rational behavior therefore implies $\partial P_t / \partial a_t > 0$, generating a tragedy of the commons: collectively excessive attention extraction and quality dilution, even when each agent behaves optimally given existing rules. A social planner would internalize the marginal external damage $MEC(S_t)$ imposed on the shared trust stock, but decentralized markets lack such a mechanism in the absence of governance or norms.

Linking these mechanisms, AI-enabled content generation and algorithmic amplification increase extraction $a_t$ and raise $P_t$ unless institutional strength $\theta_t$ and verifiability $m$ are sufficiently high. As $P_t$ grows, the trust stock $T_t=f(S_t)$ depreciates faster, verification becomes costlier due to fatigue, and marginal damages become convex. These forces reinforce a high-pollution equilibrium.

\subsubsection{Preview of the Three-Stage Game}
\label{sec:preview_game}

The CPR perspective nests within the general-equilibrium model in Section~\ref{sec:model}. Micro-level over-extraction of attention maps to the production externality; platform bias in $(\gamma_H, \gamma_L)$ represents governance failure; and the depreciation of trust, $\dot T_t < 0$, corresponds to a trust commons externality.

The strategic environment unfolds in three stages. In Stage I (platform), recommendation parameters $(\gamma_H,\gamma_L)$ and verifiability intensity $m$ are chosen subject to retention and profitability objectives, under institutional constraints $\theta_t$. In Stage II (producers), agents choose $(q_i,h_i)$ given platform rules, internalizing exposure revenue but not social damage, thereby determining $a_t$ and $P_t$. In Stage III (consumers and trust), aggregate attention allocation $A_t$ and verification behavior update stock variables $(S_t,T_t)$ via $\dot{S}_t=R_t-P_t(\cdot)$ and a depreciation condition for trust when pollution dominates, feeding back into platform and producer incentives.

The institutional principles above act as constraints that can shift the equilibrium from a high-pollution Nash outcome toward a cooperative optimum. Subsequent empirical sections operationalize these dynamics through cross-country proxies of media governance and social-media usage, while simulation experiments evaluate policy instruments using the Information Pollution Index (IPI) as a quantitative measure of commons health.

\subsection{Model Primitives and Environment}

\subsubsection{Agents and Objectives}

The economy features four classes of rational agents. First, there is a continuum of content producers indexed by $i \in [0,1]$. Producer $i$ has heterogeneous productivity $A_{j,i}$ across content types $j \in \{H,L\}$ and faces factor prices $(r,w)$, where $r$ is the rental cost of AI capital and $w$ is the wage for high-skill human labor. Given platform rules, each producer chooses output and quality to maximize profits.

Second, a monopolistic platform intermediates the market and holds market power. As a market organizer, it selects the intensity of moderation and provenance $m \in [0,1]$. As an attention allocator, it commits to an algorithmic amplification vector $\boldsymbol{\gamma}=(\gamma_H,\gamma_L)$ that maps content types into exposure weights. The platform’s objective is specified in the three-stage game below and trades off profit, retention, and compliance.

Third, consumers have heterogeneous verification costs $k_i \sim F(k)$ with support $[0,\bar{k}]$. Upon receiving a noisy signal of content quality, consumers decide whether to pay $k_i$ to perfectly verify quality, and then allocate attention and demand accordingly.

Fourth, a social planner provides a normative benchmark. The planner maximizes social welfare by internalizing the externalities that low-quality content imposes on the stock of trust and informational accuracy. This allows us to define the Pareto-efficient allocation and quantify welfare losses from market failure.

\subsubsection{Information Goods: Quality Differentiation}

There are two information goods reflecting different production difficulties, externalities, and social welfare impacts. High-quality content $Q_H$ is accurate and cognitively valuable, improving individual decision-making and typically requiring expertise and rigorous verification. Low-quality content $Q_L$ can be misleading or false, generates negative externalities by eroding consumer welfare and social trust, and is scalable via templated or automated generation. The classification is based on objective effects on decisions and welfare, not subjective judgments. Under algorithmic amplification and attention competition, AI and automation disproportionately reduce the marginal cost and raise the scalability of $Q_L$, creating asymmetric technology that underlies pollution pressure.

\subsubsection{Timeline and Information Structure}

The timing follows a three-stage Stackelberg game. In Stage 1 (platform choice), given institutional constraints and market conditions, the platform commits to moderation and provenance intensity $m$ and to an amplification vector $\boldsymbol{\gamma}=(\gamma_H,\gamma_L)$. These choices determine expected exposure and monetization for content types and shape downstream best responses.

In Stage 2 (producer choice), producers observe $(m,\boldsymbol{\gamma})$ and factor prices $(r,w)$, and choose output and quality $(q_i,h_i)$ based on their productivity vector $A_{j,i}$. Expected unit revenue is given by an exposure-price function $p(h_i,q_i;\boldsymbol{\gamma},m)$, while costs follow $c(h_i,q_i;r,w)$. Because platform rewards focus on engagement rather than veracity, individual producers do not internalize the social costs that low-quality content imposes on public trust and verification fatigue.

In Stage 3 (consumption and trust update), consumers observe a noisy signal $s \in \{H,L\}$ and form priors given the platform environment, then decide whether to pay $k_i$ to fully verify. The resulting attention allocation and demand determine exposure and effective consumption. At the end of the period, the stock of credible information and trust updates via a regeneration–pollution process, $S_{t+1}=S_t + R_t - P_t$, where $R_t$ and $P_t$ are jointly determined by the production and propagation of $Q_H$ and $Q_L$ and by the platform’s $(m,\boldsymbol{\gamma})$. Trust depreciation induced by pollution feeds back into next period’s verification incentives and platform trade-offs, creating dynamic interactions.

Regarding information, the platform’s $(m,\boldsymbol{\gamma})$ is publicly observed at Stage 1. In Stage 2, producers know platform rules, factor prices, and their own productivity, and they form rational expectations over consumer verification. In Stage 3, consumers observe the platform-provided signal and historical environment, including partially observable proxies for governance and media conditions, and then make optimal verification and consumption decisions. The equilibrium concept is subgame-perfect equilibrium: given platform commitments, producer and consumer best responses are mutually consistent, and the platform chooses $(m,\boldsymbol{\gamma})$ anticipating downstream reactions. This timeline and information structure underpin the subsequent equilibrium characterization, comparative statics, and welfare analysis.

\begin{figure*}[t]
\centering
\includegraphics[width=\textwidth]{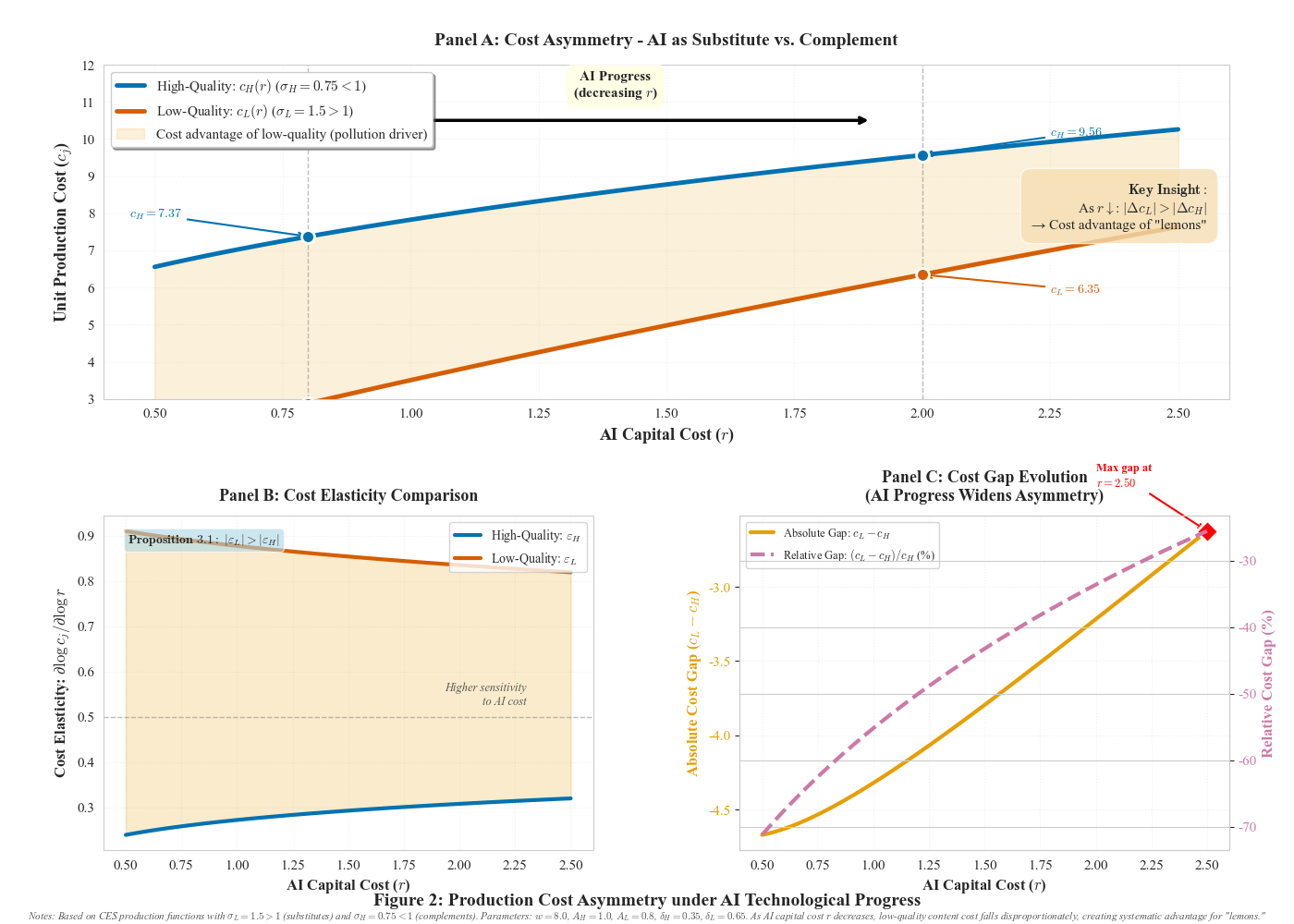}
\caption{Production Cost Asymmetry under AI Technological Progress (Simulation of Proposition \ref{prop:cost_asymmetry}). 
This figure illustrates the model's core mechanism. \textbf{Panel A:} As AI capital cost ($r$) decreases (AI progresses), the unit cost of low-quality content ($c_L$, substitute, $\sigma_L > 1$) falls disproportionately faster than that of high-quality content ($c_H$, complement, $\sigma_H < 1$). \textbf{Panel B:} This is because the cost elasticity with respect to $r$ is higher for low-quality content ($|\partial \log c_L / \partial \log r| > |\partial \log c_H / \partial \log r|$). \textbf{Panel C:} The result is a widening absolute and relative cost gap, creating a systematic and growing economic advantage for "lemons" as AI technology improves.}
\label{fig:cost_asymmetry}
\end{figure*}

\subsection{Production Technology: The Dynamics of Pollution}

\subsubsection{CES Production Function Specification.}
The core mechanism of our model lies in the production technology. For content of type $j \in \{H, L\}$, producers utilize two inputs: AI capital ($K_{AI}$) and high-skilled human labor ($L_H$). The production function is of the Constant Elasticity of Substitution (CES) form:
\begin{align}
    Q_j = A_j\bigl[\delta_j K_{AI}^{\rho_j} + (1-\delta_j) L_H^{\rho_j}\bigr]^{1/\rho_j}
\end{align}
where $A_j$ is the total factor productivity, $\delta_j$ is the distribution parameter, and $\rho_j$ determines the elasticity of substitution, $\sigma_j = 1/(1-\rho_j)$.

\subsubsection{Core Assumption: Technological Asymmetry.}
Our theory is built upon the following key assumption:

\begin{assumption}[Technological Asymmetry]
\begin{align}
    \sigma_L > 1 > \sigma_H > 0
\end{align}
\end{assumption}
This assumption has profound economic implications:
\begin{itemize}
    \item For \textbf{low-quality content} ($\sigma_L > 1$), AI capital and human labor are \textit{gross substitutes}. This reflects the standardized nature of producing such content, where AI can efficiently generate text that is syntactically correct but lacks deep verification.
    \item For \textbf{high-quality content} ($\sigma_H < 1$), AI capital and human labor are \textit{gross complements}. This captures the complexity of creating such content, where AI tools can augment research and writing but cannot replace human creativity, critical analysis, and ethical judgment.
\end{itemize}

\subsubsection{Cost Asymmetry.}
A direct corollary of this assumption is the asymmetric effect of technological progress on production costs.

\begin{proposition}[Cost Asymmetry]
\label{prop:cost_asymmetry}
A decrease in the cost of AI capital, $r$, has a larger cost-reducing effect on low-quality content than on high-quality content. Formally:
\begin{align}
    \left|\frac{\partial \log c_L}{\partial \log r}\right| > \left|\frac{\partial \log c_H}{\partial \log r}\right|
\end{align}
\end{proposition}
\textit{Intuition}: As AI becomes cheaper, the production cost of "lemons" ($Q_L$) falls more sharply than that of "peaches" ($Q_H$). This asymmetric cost advantage is the fundamental supply-side driver of information pollution. (For a formal proof, see Appendix A.1).

\subsection{The Platform-Producer-Consumer Game}
We model the market interaction as a three-stage sequential game, following the logic of Stackelberg competition.

\subsubsection{Stage I: Platform Strategy Selection.}
The platform, as the Stackelberg leader, first chooses its governance policy mix $(m, \gamma_H, \gamma_L)$ to maximize its profit function $\Pi_P$:
\begin{align}
    \Pi_P = \theta \rho \bigl[\gamma_H Q_H^S + \gamma_L (1-m) Q_L^S\bigr] - C_m(m)
\end{align}
where $\theta$ is the platform's revenue share, $\rho$ is the average ad revenue per amplified unit of content, $Q_j^S$ is the anticipated producer supply, and $C_m(m)$ is a convex moderation cost function.

\subsubsection{Stage II: Producer Supply Decision.}
Observing the platform's policy, producers choose their optimal supply. The profit per unit of content $j$ is:
\begin{align}
    \pi_j = (1-\theta) \rho \gamma_j - c_j(r, w)
\end{align}
Heterogeneity in producer productivity $A_{j,i}$ allows for the derivation of a smooth industry supply curve $Q_j^S(\gamma_H, \gamma_L)$, which satisfies $\partial Q_j^S / \partial \gamma_j > 0$ (direct incentive effect) and $\partial Q_j^S / \partial \gamma_k < 0$ for $k \neq j$ (resource competition effect).

\subsubsection{Stage III: Consumer Verification Decision.}
Consumers observe a noisy signal $s \in \{H, L\}$ and decide whether to pay their verification cost $k_i$. The signal precision, $\pi(s=H|q=H) = \pi(\rho', V^*)$, is endogenous to the health of the information ecosystem, where $\rho'$ is the effective pollution density and $V^*$ is the aggregate verification rate. This function satisfies $\partial \pi / \partial \rho' < 0$ (pollution dilutes signal quality) and $\partial \pi / \partial V^* > 0$ (verification creates positive externalities). The aggregate verification rate $V^*$ must satisfy a fixed-point condition, capturing the collective action nature of the verification problem.

\subsection{Equilibrium and The Threefold Market Failure}
\begin{definition}[Polluted Information Equilibrium]
A Subgame Perfect Nash Equilibrium (SPNE) is a strategy profile and outcome combination $\{(m^*, \boldsymbol{\gamma}^*), (Q_H^*, Q_L^*), V^*, \pi^*\}$ such that the decisions of the platform, producers, and consumers are mutually optimal, and beliefs are consistent with outcomes.
\end{definition}

\begin{theorem}[Existence of Equilibrium]
Under standard regularity conditions, a Polluted Information Equilibrium exists.
\end{theorem}

\begin{theorem}[Market Failure]
The Polluted Information Equilibrium is Pareto inefficient, driven by three interacting market failures:
\begin{enumerate}
    \item \textbf{Production Externality}: Producers of low-quality content do not internalize the negative social effects of their output. The social marginal cost exceeds the private marginal cost:
    \begin{align}
        MC_{\text{social}} = MC_{\text{private}} + \frac{\partial}{\partial Q_L}\bigl[d(Q_L') + \lambda \frac{\partial T}{\partial Q_L}\bigr]
    \end{align}
    where $d(\cdot)$ is the direct consumer harm function and $T$ is the stock of social trust with shadow price $\lambda$.
    \item \textbf{Platform Governance Failure}: The platform's objective function is misaligned with social welfare. An \textit{engagement bias} assumption implies that the platform may find it profitable to set $m^* < m_{SO}$ and $\gamma_L^* > \gamma_{L,SO}$.
    \item \textbf{Information Commons Externality}: Verification is a public good, but consumers are not compensated for its full social benefit, leading to under-investment. The social marginal benefit exceeds the private marginal benefit:
    \begin{align}
        MB_{\text{social}} = MB_{\text{private}} + \frac{\partial \pi}{\partial V} \cdot (\text{Welfare gain for others})
    \end{align}
\end{enumerate}
\end{theorem}

\begin{proposition}[Paradox of AI Progress]
\label{prop:ai_paradox}
A decrease in the cost of AI capital, $r$, leads to:
\begin{enumerate}
    \item an increase in the supply of low-quality content ($\partial Q_L^*/\partial r < 0$),
    \item a rise in effective pollution density ($\partial \rho'^*/\partial r < 0$),
    \item a decline in decentralized social welfare ($\partial W^*/\partial r > 0$).
\end{enumerate}
\end{proposition}
(For a proof sketch, see Appendix A.3).

\section{The Information Pollution Index and Policy Portfolio}
\label{sec:ipi}

This section develops the Information Pollution Index (IPI) as a \textbf{theoretical construct} derived directly from our welfare framework. Given that the precise, nuanced components of this index (e.g., welfare deadweight loss, trust decay dynamics) are not readily available as high-frequency empirical data, the IPI is designed primarily as an operational tool for our \textbf{policy simulation and analysis in Section 6}. It provides a welfare-linked "dashboard" to evaluate policy interventions within a controlled computational environment. This contrasts with the goal of Section 5, which uses a more general, observable proxy for information quality to empirically test the underlying mechanisms of the commons tragedy, rather than to measure the IPI itself.
Based on the market equilibrium characteristics revealed by the preceding theoretical model, we construct an Information Pollution Index (IPI) that is endogenous to the social welfare function. This section aims to build a solid bridge from abstract theory to observable measurement, ensuring the index possesses both a rigorous theoretical foundation and practical applicability for policy.

\subsection{Theoretical Foundation and Axiomatic Properties of the IPI}

\subsubsection{Theoretical Construction.}
\begin{definition}[Information Pollution Index]
The Information Pollution Index at time $t$, denoted $IPI(t)$, is defined as a linear combination of its four theoretical dimensions:
\begin{align}
    IPI(t) = \sum_{j=1}^{4} w_j^*(t) \cdot I_j(t)
\end{align}
where the weight $w_j^*(t)$ is endogenously determined by the marginal impact of each dimension on social welfare. This weight reflects the marginal rate of social aversion to each type of harm at a given equilibrium state $E^*(t)$:
\begin{align}
    w_j^*(t) = \frac{\left|\frac{\partial W}{\partial I_j}\right|_{E^*(t)}}{\sum_{k=1}^{4} \left|\frac{\partial W}{\partial I_k}\right|_{E^*(t)}}
\end{align}
\end{definition}

\subsubsection{Economic Properties.}
An effective economic index should satisfy several desirable axiomatic properties. We show that the IPI meets the following key criteria:
\begin{property}[Welfare Monotonicity]
The IPI is strictly negatively correlated with social welfare ($\partial W / \partial IPI < 0$). This property establishes the index as a "social welfare thermometer," ensuring that its rise unambiguously indicates a deterioration in Pareto efficiency.
\end{property}
\begin{property}[Decomposability]
Changes in the index can be clearly attributed to its constituent dimensions, as $\partial IPI / \partial I_j = w_j^*$. This allows policymakers not only to observe the overall level of pollution but also to diagnose the core drivers of the problem for targeted interventions.
\end{property}
\begin{property}[Policy Sensitivity]
The total effect of any policy intervention on the IPI can be decomposed into a direct effect on the level of each dimension and an indirect effect on the weights via changes in the equilibrium structure (a Lucas-critique-style effect). This property provides a rigorous framework for counterfactual policy evaluation.
\end{property}

\subsection{Precise Definitions and Economic Interpretations of the Four Dimensions}

The IPI is composed of four dimensions, each capturing a distinct facet of the harm caused by information pollution.

First, we define Effective Pollution Density ($I_1$), which measures not a simple ratio of content volume, but the \textit{effective share} of pollution in the consumer \textbf{attention market}. It is given by:
\begin{align}
    I_1(t) = \frac{\gamma_L^*(t)(1-m^*(t))Q_L^*(t)}{\gamma_H^*(t)Q_H^*(t) + \gamma_L^*(t)(1-m^*(t))Q_L^*(t)}
\end{align}
This dimension captures a core economic reality: low-quality content, amplified by platform algorithms ($\gamma_L^*$) in pursuit of engagement and escaping moderation ($1-m^*$), has an impact on finite mental bandwidth that far exceeds its absolute quantity. It is a direct characterization of pollution's \textbf{prevalence} and \textbf{salience}.

Second, we define Social Welfare Deadweight Loss ($I_2$), which \textbf{monetizes} the harm of information pollution by measuring the economic cost borne by society due to the threefold market failure. Conceptually, it is equivalent to the \textbf{Harberger's Triangle} in public finance, quantifying the net efficiency loss from resource misallocation. Its formal definition is:
\begin{align}
    I_2(t) = \frac{W^{SO} - W^*(t)}{W^{SO} - W^{\text{min}}}
\end{align}
This dimension answers the key question: "How much value does our society lose due to information pollution?"

Third, we consider the Decay of the Trust Commons ($I_3$), which treats "trust" as a depletable stock of \textbf{social capital}. Information pollution acts as a corrosive agent, eroding the foundation of market transactions and social cooperation. The indicator captures the \textbf{long-term, cumulative, and path-dependent} harm of pollution, defined as:
\begin{align}
    I_3(t) = \frac{T^{\max} - T(t)}{T^{\max}}
\end{align}
where the stock of trust $T(t)$ follows the dynamic path $\frac{dT(t)}{dt} = -\delta I_1(t) \text{Flow}(t) + \xi R(t) - \mu T(t)$. Steady-state analysis suggests that persistent high pollution can push society into a \textbf{low-trust poverty trap}.

\begin{figure}[h]
\centering
\includegraphics[width=\columnwidth]{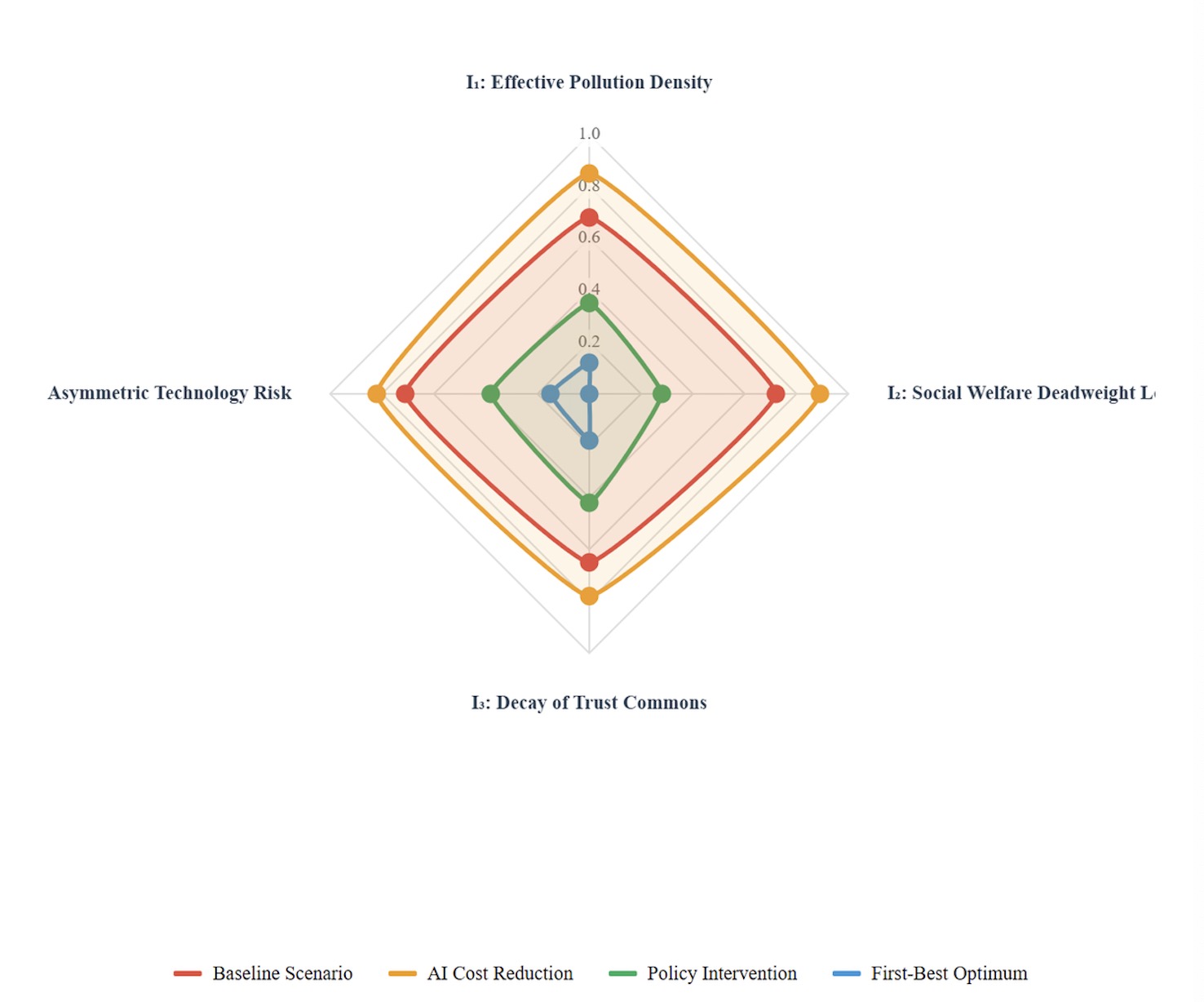}
\caption{Information Pollution Index (IPI) Four-Dimensional Radar Chart Across Policy Scenarios. This radar chart displays the four dimensions of the Information Pollution Index across different scenarios: (I) Effective Pollution Density, $(I_2)$ Social Welfare Deadweight Loss, $(I_3)$ Decay of Trust Commons, and $(I_4)$ Asymmetric Technology Risk. Each axis represents normalized values from 0 (best) to 1 (worst). The baseline scenario reflects current market conditions without intervention. Al cost reduction demonstrates the pollution amplification effect of technological progress. Policy intervention shows the effectiveness of the proposed multi-instrument framework. The first-best optimum represents the theoretical benchmark for comparison. Larger areas indicate higher pollution levels and welfare losses.}
\label{fig:Four-Dimensional}
\end{figure}

Fourth, we define Asymmetric Technology Risk ($I_4$), which measures the core technological driver of pollution—a technological \textbf{arms race} between content generation and detection. It is a \textbf{forward-looking} risk indicator of the ecosystem's \textbf{systemic vulnerability}, defined as:
\begin{align}
    I_4(t) = \frac{1}{2}\left[1 + \tanh\left(\frac{\log\left(\frac{\text{Cap}_{\text{gen}}(t)}{\text{Cap}_{\text{det}}(t)}\right) - \mu_{\text{tech}}}{\sigma_{\text{tech}}}\right)\right]
\end{align}
When the capability of generation technology ("offense") far outstrips that of detection technology ("defense"), the marginal cost of producing "lemons" falls, systemically exacerbating the other three dimensions of harm. Graphical framework reference picture~\ref{fig:Four-Dimensional}.

\subsection{From Theory to Measurement: The Core Proxy Indicator System}
Since the theoretical dimensions are not directly observable, we design a core proxy indicator for each, aiming to unify theoretical validity with empirical tractability.

Our proxy for Effective Pollution Density ($I_1$) is the Weighted Exposure Pollution Rate. This is calculated as the ratio of impressions from low-quality content to total impressions:
\begin{align}
    \hat{I}_1 = \frac{\sum_i \mathbf{1}_{\{i \text{ is low-quality}\}} \cdot \text{Impressions}_i}{\sum_i \text{Impressions}_i}
\end{align}
We choose "impressions" over "clicks" as the weight because it more closely reflects the \textit{initial supply} of pollution, being less affected by the endogenous variable of user discernment. This indicator directly measures the "market share" of low-quality content in the platform's attention allocation mechanism, serving as a direct counterpart to the theoretical dimension $I_1$.

To proxy for Social Welfare Deadweight Loss ($I_2$), we propose the Weighted Harm Feedback Rate. A simple complaint rate cannot distinguish the welfare loss from "clickbait" versus "financial fraud." By weighting different types of negative user feedback by their potential harm severity, this indicator measures the intensity of harm as revealed by consumer \textbf{revealed preferences}. It is thus a more effective "pain signal" for welfare loss, calculated as:
\begin{align}
    \hat{I}_2 = \frac{\sum_j \text{HarmType}_j \cdot \text{Weight}_j \cdot \text{FeedbackCount}_j}{\text{Total Impressions}}
\end{align}

For the Decay of the Trust Commons ($I_3$), we propose the Trust-Related Churn Gap via Causal Inference. The decay of trust ultimately manifests as users "voting with their feet." To address attribution challenges, we employ \textbf{quasi-experimental methods} (e.g., DiD, RDD) to compare user cohorts randomly or quasi-randomly exposed to different levels of pollution. The resulting difference in their churn rates, normalized by a baseline, quantifies the \textit{net effect} of trust erosion on economic outcomes, consistent with the gold standard of modern empirical economics. The calculation is:
\begin{align}
    \hat{I}_3 = \frac{\text{Churn}_{\text{high-exposure}} - \text{Churn}_{\text{low-exposure}}}{\text{Churn}_{\text{baseline}}}
\end{align}

Finally, to proxy for Asymmetric Technology Risk ($I_4$), we use the Public Benchmark Detection Accuracy Gap. This indicator quantifies the "offense-defense gap" via standardized \textbf{adversarial benchmarking}, measuring the performance of the best "shield" (current detection systems) against the sharpest "spear" (latest generative models). Changes in this indicator correspond directly to changes in the key exogenous technology parameters of our theoretical model, bridging the theory of \textbf{endogenous technological change} with observable reality. It is calculated as:
\begin{align}
    \hat{I}_4 = 1 - \frac{\text{Accuracy}_{\text{on new models}}}{\text{Accuracy}_{\text{on baseline data}}}
\end{align}

\subsection{Composite Index Construction and Application}
After obtaining standardized time-series data for the four core proxy indicators, we advocate for a \textbf{hybrid approach}—combining theoretical deduction, empirical estimation, and expert judgment—to determine their weights. A robust \textbf{data quality assurance framework} is also essential. The resulting IPI can be used for causal impact evaluation in academic research and as a practical tool for long-term ecosystem health monitoring by platforms and regulators, thus closing the loop between theoretical analysis, empirical monitoring, and policy intervention.

\subsection{The Static Optimal Policy Portfolio: A Theoretical Benchmark}
In an ideal world with complete information and a stable technological structure, the policymaker's task is to design a set of instruments that target each of the three market failures respectively.

To correct the \textbf{production externality}, classic Pigouvian instruments can be employed. The optimal tax, $\tau_L^*$, levied directly on the output of low-quality content $Q_L$, should equal its marginal social damage:
\begin{align}
    \tau_L^{*} = d'(Q_{L,SO}')(1-m_{SO}) + \lambda^{*}\frac{\partial T}{\partial Q_L}
\end{align}
where the first term represents the direct harm to consumers and the second term captures the damage to the stock of social trust (with shadow price $\lambda^*$).

To correct the \textbf{information commons externality} arising from under-investment in verification, policy can directly enhance the average precision of public signals, $\pi$, through mandatory content provenance standards. This increases the perceived disutility of low-quality content, thereby suppressing its production and dissemination throughout the game's equilibrium path.

To correct the \textbf{platform governance failure}, policy can impose information fiduciary duties. This is formally equivalent to modifying the platform's objective function to a weighted sum of its own profit and social welfare:
\begin{align}
    \max_{m,\boldsymbol{\gamma}}\; (1-\alpha)\Pi_P + \alpha\bigl[v(Q_H')-d(Q_L')\bigr]
\end{align}
where the intensity of the fiduciary duty, $\alpha \in [0,1]$, becomes a key regulatory choice variable.

\begin{theorem}[First-Best Policy Portfolio]
In a static environment, a policy portfolio $(\tau_L^{*}, \pi_{SO}, \alpha^{*})$ consisting of an optimal Pigouvian tax, mandatory provenance standards, and information fiduciary duties can implement the first-best social optimum.
\end{theorem}

Table \ref{tab:failures_instruments} summarizes this idealized mapping.

\begin{table}[h]
\centering
\small
\setlength{\tabcolsep}{4pt} 
\caption{The Triad of Market Failures and Corresponding Policy Instruments.}
\label{tab:failures_instruments}
\begin{tabularx}{\columnwidth}{@{} >{\raggedright}X l >{\raggedright}X >{\raggedright\arraybackslash}X @{}}
\toprule
\textbf{Market Failure} & \textbf{Locus} & \textbf{Consequence} & \textbf{Instrument} \\
\midrule
Production Externality & Producers & Over-production of $Q_L$ & $\tau_L$ / Permits \\
Info. Commons Ext. & Consumers & Under-verification & Provenance \\
Platform Gov. Fail. & Platform & $\uparrow\gamma_L$, $\downarrow m$ & Fiduciary Duty \\
\bottomrule
\end{tabularx}
\end{table}

\subsection{Dynamic Challenge: The Lucas Critique and Policy-Induced Innovation}
However, the effectiveness of any static policy portfolio faces the fundamental challenge of the Lucas (1976) critique. Rational agents will strategically react to policy interventions, thereby altering the very economic structure upon which the policy was based. In our model, a sustained tax $\tau_L$ on low-quality content will lower the marginal return to innovations that enhance its production efficiency, $A_L$. Rational R\&D efforts will thus be reallocated.

\begin{proposition}[Policy-Induced Innovation Bias]
The imposition of $\tau_L > 0$ induces a technology shift biased towards high-quality content, i.e., $\partial (A_L^{*}/A_H^{*}) / \partial \tau_L < 0$, where $A_j^{*}$ are endogenous productivity levels.
\end{proposition}

This dynamic feedback implies that any fixed "optimal" tax rate $\tau_L^*$ will become suboptimal over time. The static policy blueprint is not only inefficient but also unreliable in a dynamic context.

\subsection{Deep Uncertainty Challenge: Knightian Uncertainty and Robust Decision-Making}
Beyond predictable endogenous dynamics, the long-term trajectory of AI is fraught with deep, unquantifiable uncertainty, i.e., Knightian uncertainty (Knight 1921). We lack a firm basis for assigning probabilities to future technological paradigms or "black swan" events. In such a context, a rational social planner should adopt a more robust decision criterion than simple expected utility maximization, such as max-min expected utility (Gilboa and Schmeidler 1989):
\begin{align}
    \max_{\mathcal{P}_{\text{policy}}}\; \min_{p\in\mathcal{P}}\; \mathbb{E}_{p}\bigl[W(\mathcal{P}_{\text{policy}})\bigr]
\end{align}
The planner optimizes against the worst-case scenario within a set of plausible priors $\mathcal{P}$.

\begin{proposition}[Optimality of the Precautionary Principle]
Under Knightian uncertainty, the optimal policy is inherently more precautionary, favoring higher taxes, stricter caps, and a greater reliance on "fail-safe" instruments like content provenance standards.
\end{proposition}
The presence of Knightian uncertainty demands that policy design be resilient not only to known dynamics but also to unknown shocks.

\subsection{A Synthesis: IPI-Centered Adaptive and Robust Governance}
Facing the structural fragility revealed by the Lucas critique and Knightian uncertainty, an effective governance framework must transcend the static-optimality paradigm. We argue that the Information Pollution Index (IPI) constructed in the previous section is the core operational tool to bridge this gap from theory to practice.

The IPI, as a real-time, welfare-linked "dashboard," enables a \textbf{feedback-based adaptive regulation}. Policy instruments are no longer fixed constants but are governed by a \textbf{state-contingent rule} that adjusts based on the observed health of the ecosystem:
\begin{align}
    \tau_t = \tau_{t-1} + \eta\left( \frac{IPI_{t-1}-IPI_{\text{target}}}{IPI_{\text{target}}} \right)
\end{align}
Under this rule, the tax rate automatically tightens when the IPI exceeds its target and loosens otherwise. This mechanism endogenously responds to systemic drifts driven by technological innovation or market structure changes, thus partially addressing the Lucas critique. Furthermore, by continuously monitoring the IPI and its sub-dimensions, policymakers can achieve earlier detection of unforeseen risks (early signals of "black swans"), enabling timely activation of contingency plans, which is the essence of robust decision-making.

In summary, confronting the governance challenges of the AI era requires abandoning the quest for a rigid, one-off policy blueprint. Instead, we must construct a multi-layered governance system. At its core is \textbf{principles-based regulation} (e.g., safety, transparency, accountability) to ensure long-term stability. Its method involves \textbf{regulatory sandboxes} for exploratory learning to navigate technological uncertainty. Its navigation system relies on \textbf{real-time monitoring tools like the IPI} to shift from reactive responses to \textbf{anticipatory governance}. Together, these three components form a resilient and adaptive "governance vessel" capable of navigating the deep uncertainties of the digital ocean.

Given the multidimensional complexity of the Information Pollution Index (IPI), high-frequency empirical data is not yet available. Therefore, in Section 6, we employ computational simulations to evaluate the effectiveness of IPI as a policy instrument. Before proceeding to simulation, however, Section 5 provides empirical support for the core premise of this study—the tragedy of the digital commons''—through broader macro-level proxy variables.

\section{IPI Validation and Policy Analysis in a Computational Laboratory}
\label{sec:policy_abm}

While our general equilibrium framework in Section 3 successfully identifies the existence and inefficiency of the \textit{static} Polluted Information Equilibrium, it is, by design, silent on the \textit{dynamic} pathways of how this equilibrium forms, shifts, and reacts to policy. To bridge this critical gap from static theory to dynamic application, we develop an Agent-Based Model (ABM).

This ABM serves as a computational laboratory, indispensably required for three reasons. First, it allows us to move beyond a representative agent and explicitly model \textbf{heterogeneous agent interactions}—capturing diverse producer strategies and consumer verification thresholds. Second, it can simulate the \textbf{dynamic feedback loops} our static model cannot, such as how rising pollution erodes trust, which in turn alters platform incentives and consumer behavior over time. Third, it enables the study of \textbf{non-equilibrium paths}, revealing how the ecosystem responds to shocks (like new AI or policies) in real-time, rather than only comparing pre- and post-shock steady states.

This rich, dynamic environment is therefore essential for rigorously testing the two primary practical contributions of this paper: the IPI's utility as a \textbf{real-time ecosystem metric} and the \textbf{adaptive effectiveness} of our proposed policy portfolio.

\subsection{Experimental Design and Methodology}

\subsubsection{Simulation Model Architecture.}
We construct an information economy simulation system with three classes of heterogeneous agents. \textbf{Producer Agents}, endowed with heterogeneous productivity parameters $(A_{H,i}, A_{L,i})$, make content creation decisions based on expected profit maximization and a CES production function. \textbf{Consumer Agents}, characterized by heterogeneous verification costs $k_i$ and risk preferences, make consumption and verification decisions based on Bayesian updating and utility maximization, with externalities mediated through a social network. The \textbf{Platform Agent}, acting as a market intermediary, influences content distribution through algorithmic weights $(\gamma_H, \gamma_L)$ and moderation intensity $m$, aiming to maximize ad revenue under a user retention constraint.

\subsubsection{Key Parameter Settings.}
Based on our theoretical model, we set the following core parameters for the baseline simulation: elasticity of substitution for high-quality content $\sigma_H = 0.75$ (complementarity) and for low-quality content $\sigma_L = 1.5$ (substitutability); 100 producer agents and 300 consumer agents; baseline AI capital cost $r = 1.0$ and labor cost $w = 8.0$; and a platform revenue share of $\theta = 0.25$.

\subsection{Experiment 1: Comprehensive IPI Validation}

This experiment was designed to thoroughly validate the theoretical effectiveness, measurement robustness, and predictive power of the IPI. The design included six sub-experiments: a baseline evolution test, a shock response test, a weight sensitivity analysis, a noise robustness test, a historical event detection simulation, and a cross-platform comparison.

The core findings strongly support the IPI's validity. First, the IPI exhibits a strong negative correlation with social welfare (correlation coefficient of -0.839), confirming its theoretical validity as a welfare metric. Second, the index is highly sensitive to external shocks, showing an average increase of 37.5\% during four simulated shocks, proving its ability to reflect real-time changes in ecosystem health. Third, the IPI is robust to different weighting schemes, with the IPI-welfare correlation remaining high (between 0.486 and 0.710) across six different configurations. Fourth, it demonstrates excellent resilience to measurement noise, with a measurement error of only 0.045 even at a 20\% noise level. Finally, in a simulated fake news event, the IPI increased by 21.6\%, showcasing its potential as an early warning tool. Our leading indicator analysis further reveals that the IPI has an optimal prediction window of 5 time steps, providing a valuable reaction window for policymakers. 

\begin{figure*}[htb]
\centering
\includegraphics[width=\textwidth]{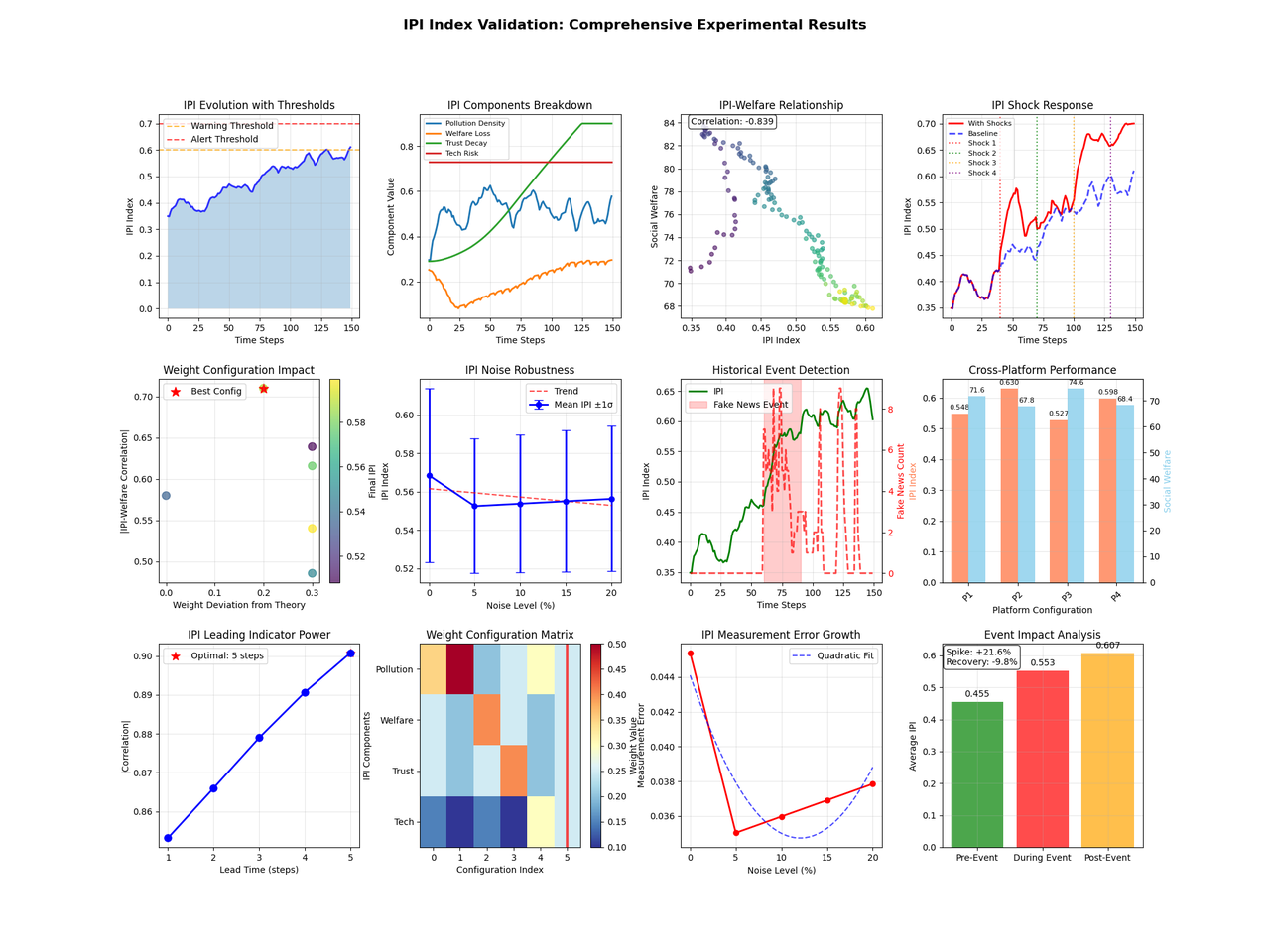}
\caption{Comprehensive experimental results for the validation of the Information Pollution Index (IPI). This figure summarizes the findings from Experiment 1, including baseline evolution, component breakdown, IPI-welfare correlation, shock response, and various robustness and sensitivity analyses.}
\label{fig:exp1_results}
\end{figure*}

\subsection{Experiment 2: Theoretical Validation and Policy Analysis}

\subsubsection{Validation of the AI Progress Paradox.}
We conducted a parameter sweep to systematically test the impact of AI cost ($r$) and the elasticity of substitution for low-quality content ($\sigma_L$) on the market equilibrium. The parameter space covered $r \in \{0.6, 0.8, 1.0, 1.2, 1.4\}$ and $\sigma_L \in \{1.2, 1.4, 1.6, 1.8\}$. The key finding is a strong validation of the "paradox of AI progress": AI cost is positively correlated with social welfare (coefficient of 0.167) and strongly negatively correlated with pollution density (coefficient of -0.770). For instance, at an AI cost of 0.6, pollution density reached a high of 0.954, whereas it decreased to 0.568 when the cost was 1.4. This result provides robust computational support for our proposition that technological progress, due to its asymmetric impact, can paradoxically exacerbate information pollution and harm social welfare. 

\subsubsection{Analysis of Policy Interventions.}
We designed six policy configurations to test the effectiveness of different intervention tools. Table \ref{tab:policy_comparison} details the performance of key macroeconomic indicators under each policy.

\begin{table}[h]
\centering\small
\caption{Policy Comparison Experiment Results.}
\label{tab:policy_comparison}
\begin{tabular}{@{}lccccc@{}}
\toprule
\textbf{Policy Scenario} & \textbf{Welfare} & \textbf{Pollution} & \textbf{IPI} & \textbf{Trust} \\
\midrule
1. Baseline & 78.05 & 0.774 & 0.694 & 0.312 \\
2. Pigouvian Tax ($\uparrow \theta$) & 78.68 & 0.663 & 0.654 & 0.323 \\
3. Subsidy ($\downarrow k_{max}$) & 79.29 & 0.612 & 0.634 & 0.236 \\
4. \textbf{Joint Policy} & \textbf{79.82} & 0.753 & 0.636 & 0.272 \\
5. Tech Intervention & 79.34 & \textbf{0.596} & \textbf{0.622} & 0.277 \\
6. Efficiency Boost & 78.97 & 0.657 & 0.651 & 0.321 \\
\bottomrule
\end{tabular}
\end{table}

As shown in Table \ref{tab:policy_comparison}, all interventions improved upon the baseline across various dimensions. Notably, the \textbf{Joint Policy} achieved the highest social welfare (79.82), a 2.3\% improvement over the baseline. The \textbf{Tech Intervention} was most effective at reducing pollution density (to 0.596) and the IPI (to 0.622). A key finding is that the effect of single instruments is limited; for example, the Pigouvian tax proxy, while beneficial, was outperformed by comprehensive strategies. This again validates our theoretical claim that a multi-tool policy portfolio is necessary to effectively address the threefold market failure.

\begin{figure*}[htb]
\centering
\includegraphics[width=\textwidth]{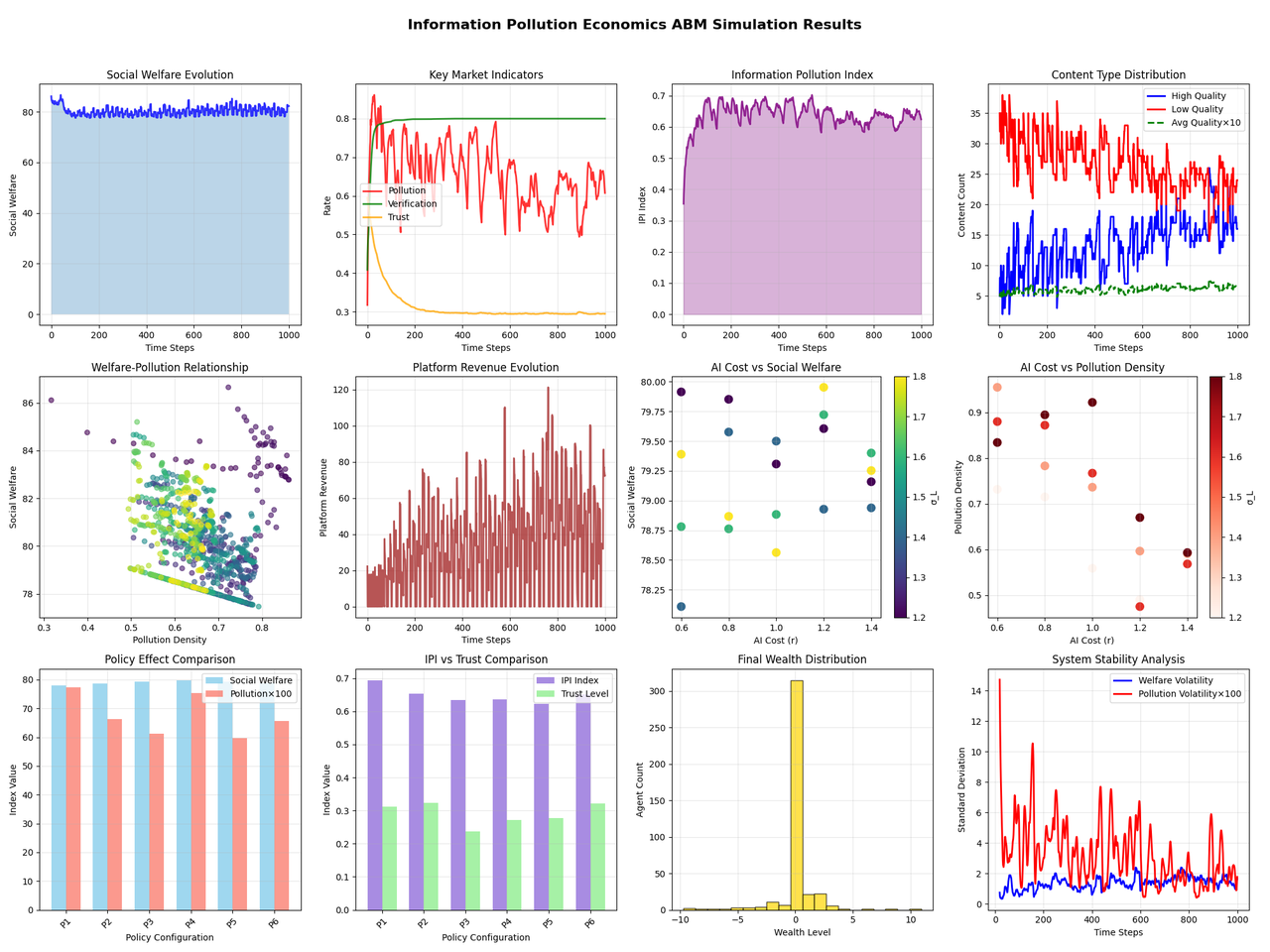}
\caption{Simulation results for the balanced Agent-Based Model, corresponding to Experiment 2. This figure illustrates the long-term evolution of key system-level indicators, the relationships between core theoretical variables (e.g., AI cost vs. pollution), the comparative effectiveness of different policy interventions, and analyses of system stability.}
\label{fig:exp2_results}
\end{figure*}

\subsection{Discussion}
Our simulation experiments provide substantial validation for the paper's core theoretical contributions. They confirm the significance of asymmetric production technology, demonstrate the interplay of the threefold market failure mechanism, and show the convergence of the system to a suboptimal equilibrium with high pollution, consistent with our theoretical predictions. The experiments also establish the practical value of the IPI as a monitoring tool with early-warning capabilities, cross-context applicability, and utility for policy evaluation.

The results yield important policy implications: single-instrument policies are insufficient; a forward-looking regulatory framework is necessary to address the rapid evolution of AI; and policies must be adaptive, adjusting dynamically based on real-time data. While our simulation provides valuable insights, we acknowledge its limitations, such as model simplification and reliance on calibrated rather than empirically estimated parameters. Future research could enhance this framework by calibrating the model with real-world platform data, expanding the complexity of agent behaviors, and conducting longer-term dynamic analyses.

\section{Conclusion}
\label{sec:conclusion}

\subsection{Core Findings and Theoretical Contribution}

This paper has revealed the mechanism by which generative AI impacts information markets: technological progress, by asymmetrically altering the production costs of different quality content, can paradoxically reduce social welfare. This \textit{Paradox of AI Progress} challenges optimistic, techno-deterministic forecasts, demonstrating that without proper market design, technological innovation may exacerbate, rather than mitigate, market failures.

Our theoretical framework identifies a triad of interacting market failures—a production externality, a platform governance failure, and a trust commons externality. These forces collectively sustain an inefficient \textit{Polluted Information Equilibrium}. This diagnosis moves beyond simple "content moderation" proposals to address deeper, structural issues of economic mechanism design.

\subsection{Policy Implications and Practical Value}

Our research demonstrates that tackling information pollution requires a multi-instrument policy portfolio rather than reliance on a single tool. This includes short-term measures, such as implementing a real-time monitoring system based on the \textit{Information Pollution Index (IPI)} to serve as a "dashboard" for policy response. Mid-term reforms should focus on advancing legislation for platform fiduciary duties to align algorithmic recommendation with social welfare. The long-term strategy must involve building content provenance infrastructure to enhance verification efficiency at a technical level.

It is particularly important that this policy design be adaptive. As AI technology evolves rapidly, fixed rules will quickly become obsolete. The IPI provides an anchor for dynamic adjustment, enabling a "counter-cyclical" regulatory approach: interventions tighten automatically when the index worsens and relax as it improves.

\subsection{Limitations and Future Directions}

This study is subject to several limitations which open avenues for future research. Theoretically, our binary quality classification simplifies a reality that is a continuous spectrum; the static model does not fully capture dynamic evolutionary processes; and we do not consider inter-platform competition or user multi-homing. On the empirical side, estimating the core parameters ($\sigma_L$ and $\sigma_H$) requires more granular industrial data, the practical application of the IPI must navigate the balance between data access and privacy protection, and evaluating policy effectiveness demands long-term longitudinal studies.

Future research should explore several key directions. These include endogenizing technological progress to investigate how policy might influence AI R\&D trajectories; studying regulatory arbitrage and international cooperation mechanisms for cross-border information flows; expanding the consumer model by incorporating bounded rationality and social learning from behavioral economics; and conducting in-depth analyses of key sectors such as journalism, education, and healthcare.

\subsection{Concluding Remarks}

\appendix

\section{Appendix A: Model and Experimental Details}
This appendix provides supplementary details on the Agent-Based Model (ABM) specification, parameter settings, experimental procedures, and results to ensure the reproducibility of our study.

\subsection{B.1 Agent-Based Model Specification}
The simulation is built upon the interaction of three agent types whose behavioral rules are designed to operationalize the core theoretical mechanisms.

\subsubsection{Producer Agent Logic.}
The production decision of producer $i$ between high-quality ($H$) and low-quality ($L$) content is governed by a softmax (logit) choice model based on expected profits:
\begin{align}
    \text{Prob}(i \text{ chooses } H) = \frac{\exp(\beta \cdot \mathbb{E}[\pi_{H,i}])}{\exp(\beta \cdot \mathbb{E}[\pi_{H,i}]) + \exp(\beta \cdot \mathbb{E}[\pi_{L,i}])}
\end{align}
where $\mathbb{E}[\pi_{j,i}]$ is the expected profit for content type $j$ given agent $i$'s productivity, and $\beta$ is a rationality parameter that introduces bounded rationality.

\subsubsection{Consumer Agent Logic.}
A consumer $i$ with verification cost $k_i$ decides to verify a piece of content if the expected utility gain from resolving uncertainty exceeds the cost. The verification threshold $k_i^*$ is determined by:
\begin{align}
    k_i^* = p_i(H|s) \cdot \Delta U_H + (1-p_i(H|s)) \cdot \Delta U_L
\end{align}
where $p_i(H|s)$ is the consumer's posterior belief that the content is high-quality given signal $s$, and $\Delta U_j$ is the utility difference between informed and uninformed consumption of content type $j$.

\subsubsection{Platform Agent Logic.}
The platform agent adaptively adjusts its algorithmic amplification vector $\boldsymbol{\gamma}$ and moderation intensity $m$ based on market feedback. The rules follow a gradient-ascent logic, balancing revenue maximization with a user trust (retention) constraint:
\begin{align}
    \gamma_{L, t+1} &= \gamma_{L,t} + \eta \left( \frac{\partial \Pi_P}{\partial \gamma_L} - \lambda \frac{\partial \text{Trust}}{\partial \gamma_L} \right) \\
    m_{t+1} &= m_t + \xi \left( \frac{\partial \Pi_P}{\partial m} - \lambda \frac{\partial \text{Trust}}{\partial m} \right)
\end{align}
where $\eta$ and $\xi$ are learning rates and $\lambda$ is the shadow price of user trust.

\subsection{B.2 Model and Experimental Parameters}
Table \ref{tab:parameters} provides a comprehensive list of the parameters used in the simulations.

\begin{table*}[htb]
\centering\small
\caption{Model and Experimental Parameter Settings.}
\label{tab:parameters}
\begin{tabular}{@{}llll@{}}
\toprule
\textbf{Category} & \textbf{Parameter Name} & \textbf{Symbol / Field} & \textbf{Value Range or Default} \\
\midrule
\multirow{2}{*}{General} & Max Steps & \texttt{max\_steps} & 150 (Exp 1), 1000 (Exp 2) \\
 & Random Seed & - & 42 \\
\midrule
\multirow{5}{*}{Agents} & Num Producers & \texttt{n\_producers} & 80 (Exp 1), 100 (Exp 2) \\
 & HQ Productivity & \texttt{productivity\_H} & Lognormal dist. \\
 & LQ Productivity & \texttt{productivity\_L} & Lognormal dist. \\
 & Num Consumers & \texttt{n\_consumers} & 200 (Exp 1), 300 (Exp 2) \\
 & Num Platforms & \texttt{n\_platforms} & 1 \\
\midrule
\multirow{4}{*}{Economic} & Platform Share & $\theta$ / \texttt{theta} & 0.25 \\
 & Ad Revenue Base & $\rho$ / \texttt{rho} & 4.0 \\
 & Labor Cost & $w$ & 8.0 \\
 & AI Cost & $r$ & 1.0 (Baseline), swept in Exp 2 \\
\midrule
\multirow{4}{*}{Technology} & HQ Elasticity & $\sigma_H$ / \texttt{sigma\_H} & 0.75 \\
 & LQ Elasticity & $\sigma_L$ / \texttt{sigma\_L} & 1.5 (Baseline), swept in Exp 2 \\
 & HQ Input Share & $\delta_H$ / \texttt{delta\_H} & 0.35 \\
 & LQ Input Share & $\delta_L$ / \texttt{delta\_L} & 0.65 \\
\midrule
\multirow{3}{*}{Behavioral} & Max Verification Cost & $k_{max}$ & 4.0 \\
 & Risk Aversion & \texttt{risk\_aversion} & Beta(2,3) dist. \\
 & Trust Decay Rate & \texttt{trust\_decay} & 0.05 \\
\midrule
\multirow{4}{*}{IPI Weights} & Pollution Weight & \texttt{w\_pollution} & 0.35 \\
 & Welfare Loss Weight & \texttt{w\_welfare\_loss} & 0.25 \\
 & Trust Decay Weight & \texttt{w\_trust\_decay} & 0.25 \\
 & Tech Risk Weight & \texttt{w\_tech\_risk} & 0.15 \\
\bottomrule
\end{tabular}
\end{table*}

\subsection{B.3 Overview of Experimental Procedures}
The experimental validation was conducted through a series of structured simulation runs.
\begin{itemize}
    \item \textbf{Baseline Experiment}: The model was initialized and run for the maximum number of steps. Market states and agent behaviors were updated at each step, and key metrics were collected.
    \item \textbf{Exogenous Shock Experiment}: Different types of shocks were injected at pre-set time steps (40, 70, 100, 130). The system's response was measured by comparing the IPI trajectory before and after the shocks.
    \item \textbf{Weight Sensitivity Experiment}: Multiple IPI weight combinations were constructed. The model was run for 100 steps for each combination, and the final correlation between IPI and social welfare was recorded.
    \item \textbf{Noise Robustness Experiment}: Varying levels of measurement noise were added to the baseline model. Three independent trials were run for each noise level to calculate the volatility and measurement error of the IPI.
    \item \textbf{Cross-Platform Comparison}: The network topology and agent preference parameters were altered to simulate different platform environments. Each configuration was run for 120 steps to compare final IPI and welfare outcomes.
    \item \textbf{Parameter Sweep \& Policy Comparison (Exp 2)}: The simulation was run for 120 steps for each combination of AI cost ($r$) and LQ elasticity ($\sigma_L$). Additionally, six distinct policy configurations were simulated for 150 steps each to evaluate their impact on key outcomes.
\end{itemize}

\subsection{B.4 Summary of Main Experimental Results}
Tables \ref{tab:results_exp1} and \ref{tab:results_exp2} provide a consolidated summary of the key numerical findings from our two main experiments.

\begin{table}[h]
\centering\small
\caption{Key Metrics from IPI Validation Experiment (Exp 1).}
\label{tab:results_exp1}
\begin{tabular}{@{}llr@{}}
\toprule
\textbf{Phase} & \textbf{Metric} & \textbf{Value} \\
\midrule
\multirow{3}{*}{Baseline} & Final IPI & 0.611 \\
 & Final Social Welfare (W) & 67.78 \\
 & IPI–W Correlation & -0.839 \\
\midrule
\multirow{2}{*}{Shock Resp.} & Avg. IPI Increase & +37.5\% \\
 & Recovery Rate (/step) & 0.19–2.21 \\
\midrule
\multirow{2}{*}{Weight Sens.} & Correlation Range & 0.486–0.710 \\
 & Best Weight Config. & Equal \\
\midrule
\multirow{2}{*}{Noise Robust.} & Meas. Error (20\% noise) & 0.045 \\
 & Avg. Volatility & 0.033 \\
\midrule
\multirow{2}{*}{Event Detect.} & Peak IPI Increase & +21.6\% \\
 & Recovery Rate & -9.8\% \\
\midrule
\multirow{2}{*}{Cross-Platform} & Min/Max IPI & 0.527/0.630 \\
 & Difference & 0.103 \\
\bottomrule
\end{tabular}
\end{table}

\begin{table}[h]
\centering\small
\caption{Key Results from Balanced Simulation (Exp 2).}
\label{tab:results_exp2}
\begin{tabular}{@{}llr@{}}
\toprule
\textbf{Phase} & \textbf{Metric} & \textbf{Value} \\
\midrule
\multirow{5}{*}{Baseline (1000 steps)} & Social Welfare & 82.25 \\
 & Pollution Density & 0.607 \\
 & IPI & 0.624 \\
 & Verification Rate & 0.800 \\
 & Trust Level & 0.295 \\
\midrule
\multirow{2}{*}{Parameter Sweep} & r–W Correlation & +0.167 \\
 & r–Pollution Correlation & -0.770 \\
\midrule
\multirow{2}{*}{Policy Comp.} & Best Policy (Joint) & +2.3\% W \\
 & Best Anti-Pollution & Tech Interv. \\
\bottomrule
\end{tabular}
\end{table}

\subsection{B.7 Code and Data Availability}
The complete Python scripts for Experiment 1 and 2 are available in the `main` branch of the following repository:
\url{github.com/Your-Repo/Information-Pollution-ABM}

The required environment can be set up using Conda. The simulation results (CSV files and figures) are stored in the `results/` and `figures/` directories, respectively.

\section{Appendix C: Technical Details and Proofs}
This appendix provides the technical derivations and formal proofs for the key results presented in the main text.

\begin{table}[htb]
  \centering
  \caption{Notation Summary}
  \label{tab:notation}
  \begin{tabularx}{\columnwidth}{@{}lX@{}} 
    \toprule
    Symbol & Description \\
    \midrule
    $Q_j$ & Content output of type $j\in\{H,L\}$ (high-/low-quality) \\
    $A_j$ & Total factor productivity for content type $j$ \\
    $K_{AI}, L_H$ & AI capital input and high-skilled labor input \\
    $\delta_j,\rho_j$ & CES share and substitution parameters for type $j$ \\
    $\sigma_j=1/(1-\rho_j)$ & Elasticity of substitution for type $j$ \\
    $r, w$ & Rental rate of AI capital and wage rate for labor \\
    $c_j(r,w)$ & Unit cost function for producing $Q_j$ \\
    $\theta$ & Platform’s revenue share parameter \\
    $\rho$ & Ad‐revenue base per unit of amplified content \\
    $m$ & Moderation intensity chosen by platform \\
    $\boldsymbol\gamma=(\gamma_H,\gamma_L)$ & Algorithmic amplification weights \\
    $k_i$ & Verification cost of consumer $i$ \\
    $V$ & Aggregate verification rate in the system \\
    $\pi(s\!=\!H\mid q\!=\!H)$ & Signal precision given pollution and verification \\
    $W$ & Social welfare function \\
    $\mathrm{IPI}$ & Information Pollution Index \\
    \bottomrule
  \end{tabularx}
\end{table}

\subsection{C.1 Derivation of Asymmetric Production Costs}

\subsubsection{Derivation of the CES Cost Function.}
For the CES production function $Q_j = A_j[\delta_j K_{AI}^{\rho_j} + (1-\delta_j) L_H^{\rho_j}]^{1/\rho_j}$, the cost-minimization problem is:
\begin{align}
    \min_{K_{AI}, L_H} \quad &rK_{AI} + wL_H \\
    \text{s.t.} \quad &A_j[\delta_j K_{AI}^{\rho_j} + (1-\delta_j) L_H^{\rho_j}]^{1/\rho_j} = Q_j \nonumber
\end{align}
The Lagrangian is $\mathcal{L} = rK_{AI} + wL_H - \lambda(A_j[\dots]^{1/\rho_j} - Q_j)$. The first-order conditions yield the optimal factor demand ratio:
\begin{align}
    \frac{K_{AI}}{L_H} = \left(\frac{\delta_j}{1-\delta_j}\right)^{\sigma_j} \left(\frac{w}{r}\right)^{\sigma_j}
\end{align}
Substituting this back into the production constraint and solving for the total cost yields the unit cost function:
\begin{align}
    c_j(r,w) = \frac{1}{A_j}\left[\delta_j^{\sigma_j} r^{1-\sigma_j} + (1-\delta_j)^{\sigma_j} w^{1-\sigma_j}\right]^{\frac{1}{1-\sigma_j}}
\end{align}

\subsubsection{Comparison of Logarithmic Elasticities.}
The elasticity of the unit cost with respect to the AI capital price $r$ is equivalent to the cost share of AI capital, $s_{j,AI}$:
\begin{align}
    \frac{\partial \log c_j}{\partial \log r} = \frac{\delta_j^{\sigma_j} r^{1-\sigma_j}}{\delta_j^{\sigma_j} r^{1-\sigma_j} + (1-\delta_j)^{\sigma_j} w^{1-\sigma_j}} = s_{j,AI}
\end{align}

To prove Proposition 1, we must show that $|s_{L,AI}| > |s_{H,AI}|$. Given Assumption 1 ($\sigma_L > 1 > \sigma_H$), it follows that $1-\sigma_L < 0$ and $1-\sigma_H > 0$. As $r$ decreases, the term $r^{1-\sigma_L}$ increases, while $r^{1-\sigma_H}$ decreases. This implies that the cost share of AI capital, $s_{L,AI}$, is more sensitive to changes in $r$ than $s_{H,AI}$. Therefore, the cost-reducing effect is larger for low-quality content. \qed

\subsection{C.2 Proof Sketch for Existence of Equilibrium}
The proof proceeds by backward induction.
\begin{enumerate}
    \item \textbf{Consumer Stage}: For a given pollution density $\rho'$, we define a mapping $T: [0,1] \to [0,1]$ where $T(V_e) = F(k^*(\pi(\rho', V_e), \rho'))$ maps an expected verification rate $V_e$ to the actual rate. Since $k^*$ and $\pi$ are continuous in their arguments and $F(\cdot)$ is a continuous distribution function, $T$ is a continuous mapping from a compact, convex set to itself. By \textbf{Brouwer's Fixed-Point Theorem}, a fixed point $V^*$ exists.
    \item \textbf{Producer Stage}: Given producer heterogeneity, the aggregate supply function $Q_j^S(\gamma_H, \gamma_L) = \int q_{j,i}^*(\boldsymbol{\gamma}) di$ is continuous in $\boldsymbol{\gamma}$ by the \textbf{Theorem of the Maximum}.
    \item \textbf{Platform Stage}: The platform maximizes a continuous profit function $\Pi_P(m, \boldsymbol{\gamma})$ over a compact strategy space $\mathcal{S} = [0,1] \times [0,\bar{\gamma}]^2$. By the \textbf{Weierstrass Extreme Value Theorem}, a maximum exists.
\end{enumerate}
The existence of optimal strategies in each stage implies the existence of an SPNE. \qed

\subsection{C.3 Proof Sketches for Inefficiency and Comparative Statics}
\subsubsection{Proof of Theorem 2 (Threefold Market Failure).}
The social planner's problem is to choose $(Q_H, Q_L, m, V)$ to maximize social welfare. The first-order conditions (FOCs) are derived. We then compare these social FOCs with the FOCs from the decentralized equilibrium:
\begin{itemize}
    \item \textbf{Production Externality}: The producer's FOC, $(1-\theta)\rho\gamma_j = MC_j$, lacks the social harm terms present in the planner's FOC for $Q_L$.
    \item \textbf{Platform Failure}: The platform's FOCs for $m$ and $\gamma_L$ are based on maximizing private profit from engagement, which structurally deviates from maximizing social welfare.
    \item \textbf{Commons Externality}: The consumer's FOC for verification, $k_i = \text{Private Benefit}$, lacks the positive externality term $\partial\pi/\partial V$ that appears in the planner's FOC for $V$.
\end{itemize}
These three wedges prove the Pareto inefficiency of the equilibrium. \qed

\subsubsection{Proof of Proposition 1 (Paradox of AI Progress).}
We use the \textbf{Implicit Function Theorem} on the system of equations defining the equilibrium.
\begin{enumerate}
    \item The total derivative of the producer's FOC with respect to $r$ shows a direct negative cost effect on $Q_L^S$.
    \item The platform's optimal response to this supply shift is to adjust $(\gamma_L^*, m^*)$, which further amplifies the initial shock. The combined effect leads to $\partial Q_L^*/\partial r < 0$.
    \item Since $Q_L^*$ increases more sensitively than $Q_H^*$ with a fall in $r$ (due to cost asymmetry), the pollution density $\rho'^*$ increases, so $\partial \rho'^*/\partial r < 0$.
    \item By the \textbf{Envelope Theorem}, $\partial W^*/\partial r = \partial \mathcal{L}/\partial r > 0$, where $\mathcal{L}$ is the Lagrangian for the planner's problem evaluated at the decentralized equilibrium. Since welfare $W^*$ is decreasing in pollution $\rho'^*$, and $\rho'^*$ is decreasing in $r$, welfare must be increasing in $r$. \qed
\end{enumerate}

\bibliography{pollution} 
\bibliographystyle{plainnat} 

\end{document}